\documentclass[letterpaper,twocolumn,10pt]{article}
\usepackage{usenix}
\usepackage{multirow}
\usepackage{booktabs}
\usepackage{graphicx}

\usepackage{tikz}
\usepackage{amsmath}

\usepackage{filecontents}

\usepackage{paralist}
\usepackage{booktabs}
\usepackage{multirow}
\usepackage{listings}
\usepackage{subcaption}
\usepackage{tikz}
\usepackage{xcolor}
\usepackage{pifont}
\usepackage{amsmath}
\usepackage{amsthm}
\theoremstyle{definition}
\usepackage{footnote}

\usepackage{enumitem}
\usepackage[normalem]{ulem}

\usepackage{enumitem}
\setitemize{noitemsep,topsep=0pt,parsep=0pt,partopsep=0pt}

\makesavenoteenv{tabular}
\makesavenoteenv{table}

\newcommand{\Romannum}[1]{\uppercase\expandafter{\romannumeral #1\relax}}

\newcommand{\ncases}{$26$}
\newcommand{\ncves}{$52$}

\def\WithComments{}
\ifdefined \WithComments
\newcommand{\zhc}[1]{\textcolor{blue}{\textbf{#1}}}
\else
\newcommand{\zhc}[1]{}
\fi

\usepackage{tikz}
\usepackage{amsmath}
\usetikzlibrary{shapes.misc,shadows}

\usepackage[compact]{titlesec}
\usepackage{microtype}
\def\BibTeX{{\rm B\kern-.05em{\sc i\kern-.025em b}\kern-.08em
    T\kern-.1667em\lower.7ex\hbox{E}\kern-.125emX}}

\usepackage{tikz}
\usepackage{xcolor}

\usepackage{listings}

\definecolor{codegreen}{rgb}{0,0.6,0}
\definecolor{codegray}{rgb}{0.5,0.5,0.5}
\definecolor{codepurple}{rgb}{0.58,0,0.82}
\definecolor{backcolour}{rgb}{0.95,0.95,0.92}

\lstdefinestyle{mystyle}{
    backgroundcolor=\color{backcolour},
    commentstyle=\color{codegreen},
    keywordstyle=\color{magenta},
    numberstyle=\tiny\color{codegray},
    stringstyle=\color{codepurple},
    basicstyle=\ttfamily\footnotesize,
    breakatwhitespace=false,
    breaklines=true,
    captionpos=b,
    keepspaces=true,
    numbers=left,
    numbersep=5pt,
    showspaces=false,
    showstringspaces=false,
    showtabs=false,
    tabsize=2
}

\usepackage{filecontents}

\begin{document}

\date{}

\title{\Large \bf 
Characteristics, Root Causes, and Detection of\\
  Incomplete Security Bug Fixes in the Linux Kernel}

\author{
Qiang Liu$^{1}$\thanks{All work was done by Aug., 2022.},
Wenlong Zhang$^{1}$,
Muhui Jiang$^{2,1}$,
Lei Wu$^{1}$,
Yajin Zhou$^{1}$\\
$^{1}$Zhejiang University,
$^{2}$The Hong Kong Polytechnic University 
}

\maketitle

\begin{abstract}

Security bugs in the Linux kernel emerge endlessly and have attracted much
attention.
However, fixing security bugs in the Linux kernel could be incomplete due to
human mistakes.
Specifically, an incomplete fix fails to repair all the original security
defects in the software, fails to properly repair the original security defects,
or introduces new ones.

In this paper, we study the fixes of incomplete security bugs in the Linux
kernel for the first time, and reveal their characteristics, root causes as well
as detection.
We first construct a dataset of incomplete security bug fixes in the Linux
kernel and answer the following three questions.
What are the characteristics of incomplete security bug fixes in the Linux
kernel? What are the root causes behind them? How should they be detected to
reduce security risks?
We then have the three main insights in the following.

\textbf{Insight 1:} The fixes of incomplete security bugs in the Linux kernel
are gradually decreasing, and happened occasionally recently; their incubation
period is 2 to 7 months; they are more likely to appear when denial of service
vulnerabilities need to be fixed; the number of files and code changes affected
by related patches is generally small.

\textbf{Insight 2:} There are three common root causes behind the incomplete
security bug fixes in the Linux kernel. First, a developer was misled by
superficial phenomena of the crash before the fix; second, a developer did not
pay attention to the similar modules during the fix; third, a developer
introduced a semantic error after the fix.

\textbf{Insight 3:} Aiming at the root cause of ``not paying attention to
similar modules when fixing'', we propose a fix-pattern-based detection
algorithm.
Using this algorithm, we successfully find a new case of and a hidden case of
incomplete security bug fixes in the Linux kernel.
Compared with most of the previous empirical research that stays in suggestion,
this paper has a certain degree of advancement.

\end{abstract}

\section{Introduction}

Linux kernel, as a critical component of both the open-source software supply
chain and enterprise core software, has long been a prime target for attackers
and faces substantial security risks.
To date, more than 2,650 security flaws with assigned CVE identifiers have been
discovered in the Linux kernel~\cite{linux-cvedetails}.
These vulnerabilities inevitably impose varying degrees of security and
operational risks on downstream open-source software, end users relying on the
Linux kernel, and even enterprises and government agencies.

Security flaws in open-source software emerge continuously, and stakeholders
expect these flaws to be correctly fixed.
However, studies have shown that \emph{incomplete security bug fixes} account
for 12\% of all security fixes, respectively~\cite{r1-v-ccs17}, exerting a
non-negligible impact on the security of open-source ecosystems.
An incomplete security bug fix refers to a situation where not all code affected
by a security flaw is fully patched, thereby leaving the original flaw partially
or entirely unfixed, or the patch itself contains mistakes, either failing to
remediate the original flaw or introducing a new one. 
Evidently, the existence of such incomplete fixes continues to expose
open-source software to potential exploitation.
Understanding their characteristics, root causes, and effective prevention
measures is therefore of significant value to both researchers and software
maintainers.

To reveal the characteristics, underlying causes, and detection strategies of
incomplete security bug fixes in the Linux kernel, this paper constructs the
first dataset dedicated to this phenomenon. 
First, using quantitative research methods, we identify and characterize
patterns exhibited by incomplete fixes.
Although security bug fixing in the Linux kernel follows stricter processes than
general bug fixing, our findings demonstrate that developers involved in fixing
security flaws still make mistakes.
Next, through an empirical study, we analyze for the first time the root causes
of \ncases{} incomplete security bug fixes in the Linux kernel from the
perspectives of pre-fix, in-fix, and post-fix activities, and provide
improvement suggestions for the overall security bug fixing workflow. 
Finally, we conduct an in-depth investigation of a prevalent category of
incomplete fixes, those arising from overlooking analogous modules during patch
development.
By leveraging FixMiner~\cite{anil2020fixminer}, a semantic-aware
syntax-tree–based similarity analysis tool, we identify two previously unknown
incomplete security bug fixes in the Linux kernel.
This tool can assist in detecting this specific type of incomplete security fix
and serves as a complementary approach to traditional Linux kernel regression
testing.

Although prior empirical studies have examined incomplete bug
fixes~\cite{yin2011fixes,r1-c-msr12,an2014supplementary,mi2016empirical,ni2020analyzing},
this work differs in both content and practical contributions in two key aspects
(see Section~\ref{sec:related-work}).
First, existing studies primarily focus on non-system software such as Eclipse
and Mozilla, whereas our study targets the Linux kernel, a representative and
complex system software.
Second, previous empirical studies often stop at providing statistical summaries
or qualitative recommendations. 
In contrast, we go beyond characterization and, based on our empirical findings,
leverage fix-pattern fingerprints to discover two previously unreported
incomplete security bug fixes in the Linux kernel (both have been fixed).

\textbf{Contributions.}
This paper has the following contributions:

\begin{itemize}
\item First dataset of incomplete security bug fixes in the Linux kernel at
\url{https://doi.org/10.5281/zenodo.6423844},
enabling systematic study of this long-overlooked phenomenon.

\item Empirical characterization of incomplete security bug fixes. Through
quantitative and qualitative analysis of \ncases{} cases, we reveal their
temporal trends, vulnerability types, patch characteristics, and the recurring
root causes across pre-fix, in-fix, and post-fix stages.

\item Fix-pattern–based detection method.
We design a detection algorithm based on fix-pattern fingerprinting using
FixMiner.
Applying it to 1,615 Linux kernel CVEs uncovers two previously unknown
incomplete security fixes, demonstrating the method’s effectiveness.

\end{itemize}

\section{Dataset and Research Questions}\label{sec:methodology}

This section describes how we construct the dataset of incomplete security bug
fixes in the Linux kernel and the three research questions on the
characteristics, root causes, and detection of incomplete security bug fixes.

\begin{table*}[!t]
\centering
\caption{Bug-tracking platforms and vulnerability databases}
\label{tab:db-list}
\footnotesize
\begin{tabular}{@{}ccc@{}}
\toprule
Platform/Database & Link & Description \\
\midrule
Red Hat Bugzilla & \url{https://bugzilla.redhat.com/query.cgi?format=advanced} & Red Hat bug tracker \\
CVE Mitre & \url{https://cve.mitre.org/cve/search_cve_list.html} & CVE archival database \\
NDV & \url{https://nvd.nist.gov/vuln/full-listing} & National Vulnerability Database \\
CERT/CC VND & \url{https://www.kb.cert.org/vuls/search/} & CMU vulnerability notes database \\
LWN & \url{https://lwn.net/Security/Index/} & Linux kernel news site \\
oss-sec & \url{https://seclists.org/oss-sec/} & Open-source security mailing list \\
fulldisc & \url{https://seclists.org/fulldisclosure/} & Full-disclosure mailing list \\
bugtraq & \url{https://seclists.org/bugtraq/} & Bugtraq mailing list \\
Exploit-DB & \url{https://www.exploit-db.com/search} & Exploit database \\
VED & \url{https://www.rapid7.com/db/} & Vulnerability/exploit database \\
Ubuntu & \url{https://ubuntu.com/security/cve} & Ubuntu CVE reports \\
Gentoo & \url{https://security.gentoo.org/glsa} & Gentoo security database \\
SUSE Bugzilla & \url{https://bugzilla.suse.com/query.cgi?format=advanced} & SUSE bug tracker \\
SUSE CVE & \url{https://www.suse.com/security/cve/} & SUSE CVE database \\
Debian & \url{https://security-tracker.debian.org/tracker/data/json} & Debian CVE dataset \\
CVE details & \url{https://www.cvedetails.com/browse-by-date.php} & Public vulnerability datasource \\
Other vendors & \url{https://oss-security.openwall.org/wiki/vendors} & Security vendor list \\
\bottomrule
\end{tabular}%
\end{table*}

\subsection{Dataset}

The dataset of incomplete security bug fixes refers to a collection of patch
sets in which the initial fix of a security flaw is either incomplete.
Each entry in this dataset contains two or more patches.
The first patch corresponds to the initial fix of a security vulnerability
(denoted as Fix-0), and any subsequent patches (Fix-1, Fix-2, and so on) further
correct the incompleteness or mistakes in Fix-0.
Each patch consists of two components: a textual description of the
vulnerability being addressed and the repair strategy used, and the
corresponding patch code.
In addition, we enrich each entry with the vulnerability type using the
classification scheme published by CVE Details (``The ultimate security
vulnerability datasource'')~\cite{type-cvedetails}.
Finally, each dataset entry is indexed by the CVE (Common Vulnerabilities \&
Exposures) identifier associated with Fix-0.

Constructing a comprehensive dataset of incomplete security bug fixes is highly
challenging.
For instance, Li et~al.~\cite{r1-v-ccs17}, in their study of security patches,
identified fixes by extracting patch links listed in the reference section of
each CVE entry, which is easy to automate.
However, it is unsuitable for identifying incomplete security fixes, because
real-world CVE reference links only include the patch that claims to fix the
vulnerability, and never include information about subsequent corrective
patches.
Information (such as \textit{incomplete} or \textit{incorrect}) is typically
found in the textual descriptions of the vulnerability or the surrounding
discussion.
Furthermore, not all Linux kernel security fixes are assigned CVE identifiers,
and identifying which patches are security, related remains a difficult problem.
Although machine learning–based techniques exist~\cite{yaqin2021labelcommits},
they still require manual inspection and labeling by security experts for
training and evaluation. 
To date, no public dataset or automated tool exists for determining whether a
Linux kernel patch is security-related.

In this work, we retrieved data from 17 bug-tracking platforms and vulnerability
databases and ultimately collected \ncases{} incomplete security bug fix
instances from the Linux kernel, covering \ncves{} CVEs.
Our approach uses two complementary search strategies.
First, we performed a forward search on the CVE Mitre platform using the
keywords \textit{incomplete}, \textit{incorrect}, and \textit{Linux kernel}.
Second, we conducted a reverse search by querying only \textit{incomplete} and
\textit{incorrect} on the CVE Mitre platform, obtaining more than 300 results
across various projects and manually filtering those affecting the Linux kernel.
We then applied the same strategies across the remaining 16 platforms and
databases listed in Table~\ref{tab:db-list}.
Through manual consolidation and deduplication, we obtained \ncases{} incomplete
security bug fix instances and further enriched them with patch descriptions,
patch code, and vulnerability type information.\footnote{CVE-2012-3552 is also
incomplete, but its patch is unavailable and therefore excluded.}
The dataset is publicly available at
\url{https://doi.org/10.5281/zenodo.6423844} for the research community.

\begin{figure}[!t]
\centering
\includegraphics[width=\linewidth]{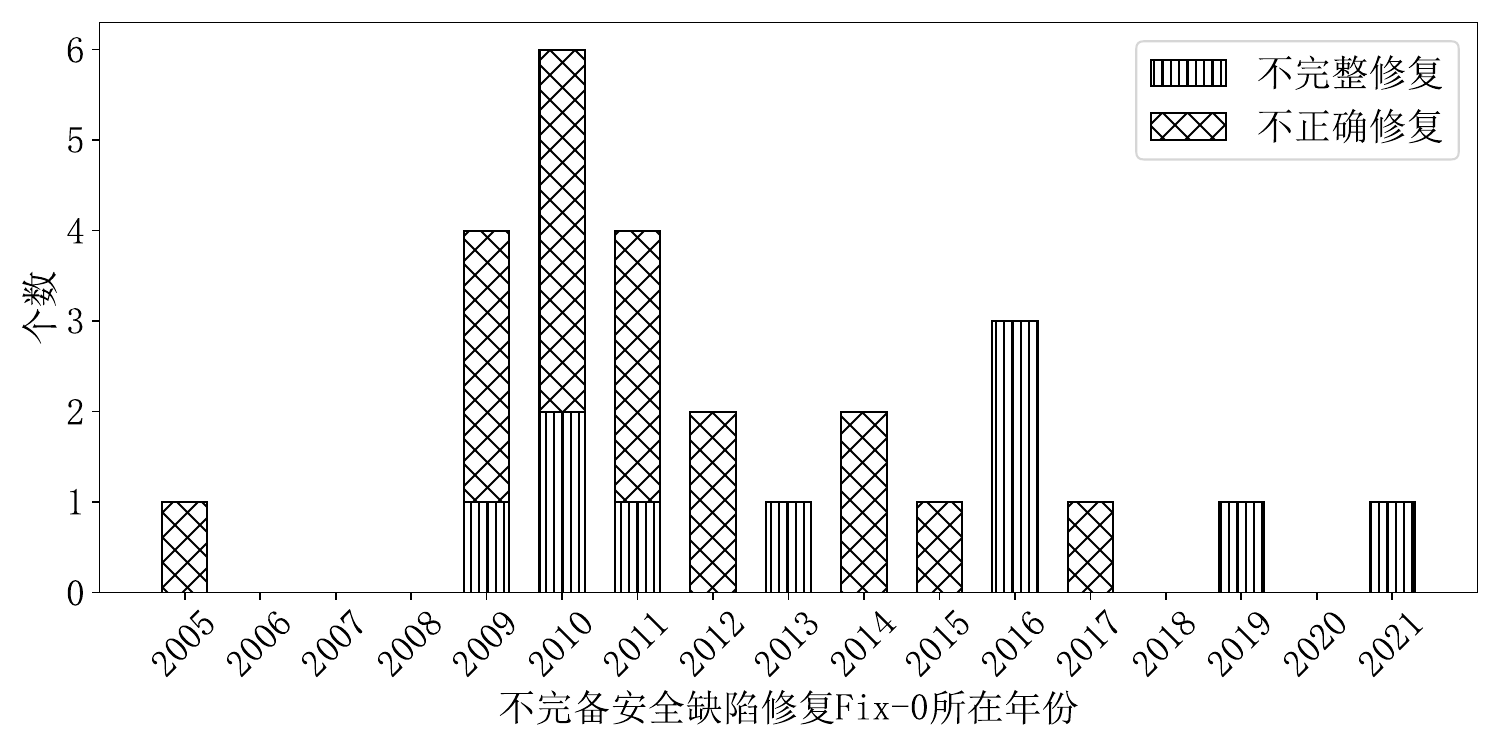} 
\caption{Annual trend of incomplete security bug fixes}
\label{img2} 
\end{figure}

\begin{figure}[!t]
\centering
\includegraphics[width=\linewidth]{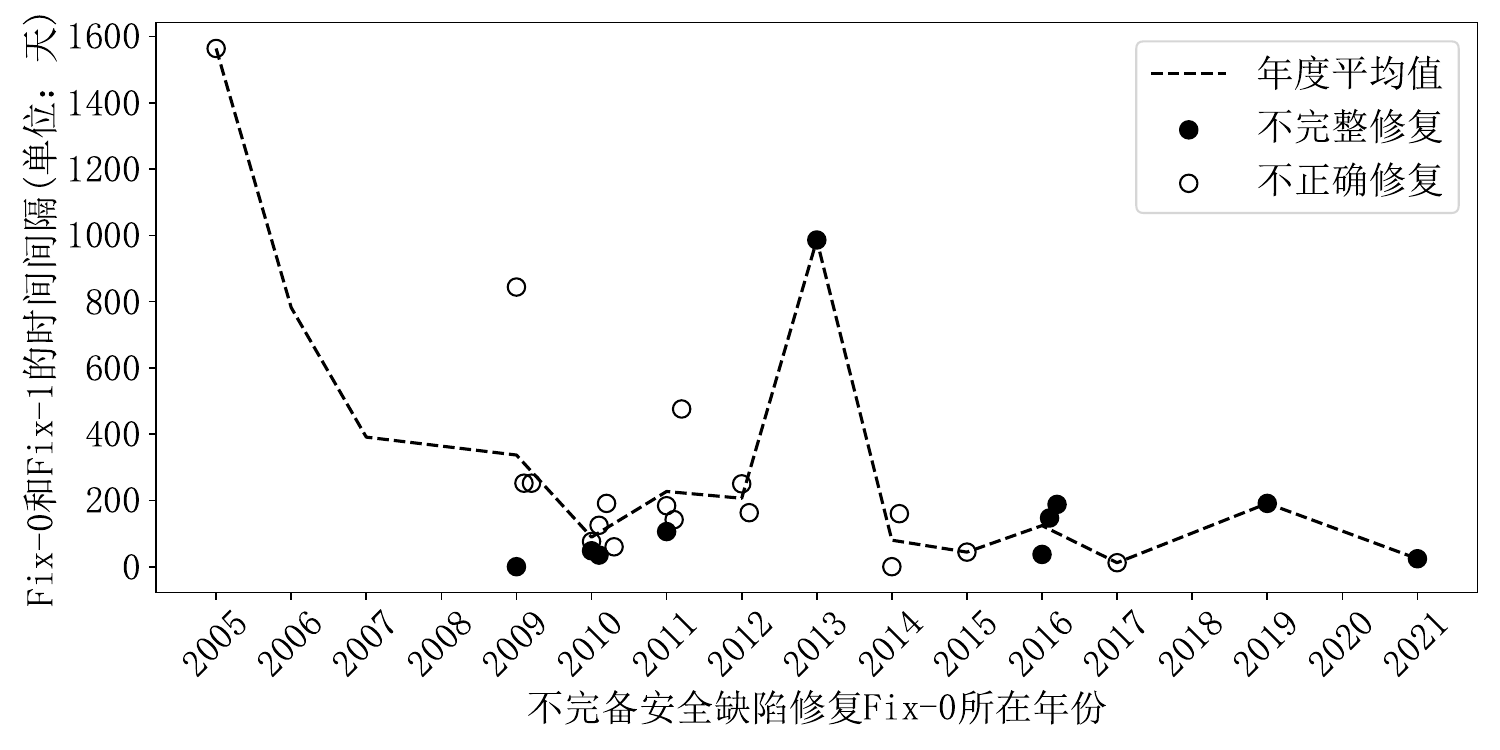} 
\caption{Time interval between discovery and initial fix}
\label{img3} 
\end{figure}

\subsection{Research Questions}

Based on the constructed dataset of incomplete security bug fixes, we
investigate the following three research questions, progressing from
phenomenon-level characterization to causal analysis and finally to detection.

\begin{itemize}

\item \textbf{Research Question 1: What are the characteristics of incomplete
security bug fixes?}
We first study the temporal distribution of incomplete security bug fix
instances in the Linux kernel to understand the phenomenon longitudinally
(RQ1.1).
We then examine the time lag between Fix-0 and subsequent fixes, which reflects
the degree of stealthiness of incomplete fixes (RQ1.2).
Next, we analyze the relationship between incomplete fixes and vulnerability
types to inform resource allocation for different categories of security flaws
(RQ1.3).
Finally, we investigate the characteristics of the patch files themselves to lay
the groundwork for identifying the root causes of these incomplete fixes
(RQ1.4).

\item \textbf{Research Question 2: What are the root causes of incomplete
security bug fixes?}
We analyze and summarize the causes of incomplete fixes in chronological order,
before the initial fix, during the fixing process, and after the fix is
released, to provide guidance for developing more complete security fixes and to
inform the design of detection techniques.

\item \textbf{Research Question 3: How can incomplete security bug fixes be
detected?}
Finally, we develop a detection tool inspired by the identified root causes and
apply it to the Linux kernel to discover additional incomplete security fixes,
thereby reducing the likelihood of such fixes in the future.

\end{itemize}

\section{RQ1: What Are the Characteristics of Incomplete Security Bug Fixes?}
\label{sec:characteristics}

In this section, we conduct a quantitative analysis of incomplete security bug
fixes in the Linux kernel and answer each subquestion of RQ1.

\subsection{RQ1.1: Temporal Distribution of Incomplete Security Bug Fixes}

As shown in Figure~\ref{img2}, we categorize incomplete security bug fixes in
the Linux kernel by year. 
Based on our definitions and manual inspection, 11 cases fail to repair all the
original security defects in the software, and 15 cases fail to properly repair
the original security defects, or introduces new ones.

\begin{figure*}[!t]
\centering
\includegraphics[width=.6\textwidth]{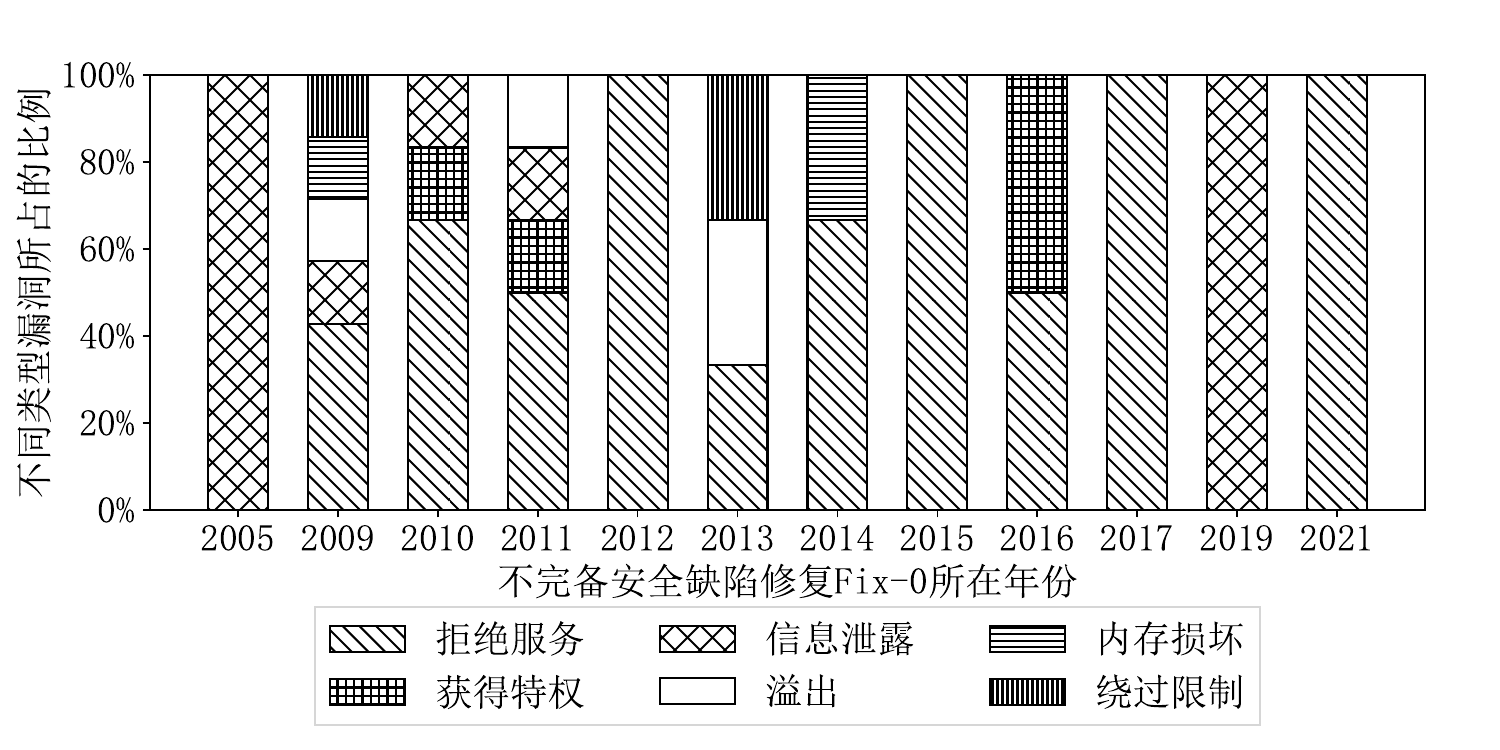}
\caption{Vulnerability type distribution of Fix-0 in incomplete security bug fixes}
\label{fig:vul-type}
\end{figure*}

\begin{table*}[t]
\centering
\footnotesize
\begin{tabular}{c|ccccc}
\toprule
Dataset   & Modified Files & Added LOC & Deleted LOC & Modified LOC & Patch Size (bytes) \\ \midrule
Fix-0 & 2.32/3.21   & 32.76/58.23  & 11.04/24.32 & 9.50/14.54 &3173.68/4452.91 \\
Fix-1 & 2.08/2.27   & 26.20/49.52  &  9.92/21.61 & 367/3.29  &2456.96/3992.47 \\ 
\bottomrule
\end{tabular}
\caption{Patch characteristics of incomplete security fixes (mean/std)}
\label{tab:patch-stats}

\end{table*}

\textbf{Finding 1:}
Descriptions of incomplete security fixes commonly use the keywords
\textit{incomplete} and \textit{incorrect}. 
However, manual inspection reveals misuses of the term \textit{incomplete}. 
For instance, the fix for CVE-2010-4163 is an incorrect fix, but its public
description labels it as incomplete. 
Such inconsistencies can mislead subsequent analysis.
Therefore, we publish our curated dataset with corrected descriptions that
strictly follow our definitions.

\textbf{Finding 2:}
As shown in Figure~\ref{img2}, incomplete security bug fixes appeared as early
as 2005.  Over time, their annual counts show a general downward trend starting
from 2010, though the phenomenon remains non-negligible.  A short-term rebound
appears around 2016, likely correlated with the spike of 215 and 449 Linux
kernel vulnerabilities reported in 2016 and 2017, respectively.  Overall,
publicly available data indicate that while occasional increases occur,
incomplete fixes show a gradual decline, reflecting more careful security
patching practices.  Nevertheless, incomplete fixes still appeared in 2021, and
this study also discovered new instances (see Section~\ref{sec:detection}).  
These observations confirm that incomplete security bug fixes remain an
important issue.

\subsection{RQ1.2: Time Interval Between Initial Fix and Subsequent Correction}

Figure~\ref{img3} shows the time interval (in days) between Fix-0 and Fix-1 for
each incomplete security bug fix, showing how long it took for the incomplete
fix to be discovered.

\textbf{Finding 3:}
Although the first observed case of an incomplete fix (in 2005) took nearly four
years (over 1,600 days) to be corrected,  Figure~\ref{img3} shows that the Fix-0
to Fix-1 interval has gradually shortened over time.  This suggests incomplete
fixes have become easier to uncover and are now detected more promptly,
indicating decreasing stealthiness.

\textbf{Finding 4:}
Most incomplete security bug fixes are detected within 50--200 days. 
In some cases, Fix-1 was committed on the same day as Fix-0.
However, some fixes remained hidden for years.
For example, the spelling error introduced during the fix for CVE-2005-4881 was
not discovered until 1,564 days later, leading to an uninitialized data leak
(CVE-2009-3612). 
These cases imply that additional yet-undiscovered incomplete fixes may still
exist among historical patches.

\subsection{RQ1.3: Relationship Between Incomplete Fixes and Vulnerability
Types}

As shown in Figure~\ref{fig:vul-type}, we categorize Fix-0 patches by the
vulnerability types defined by CVE Details.

\textbf{Finding 5:}
Since 2009, the diversity of vulnerability types associated with incomplete
fixes has gradually decreased. 
According to CVE Details classifications, denial-of-service vulnerabilities
dominate, followed by privilege escalation, and information disclosure.
Therefore, when fixing these high-frequency categories, additional reviewers,
testers, and discussion efforts should be allocated to avoid incomplete fixes.

\subsection{RQ1.4: Patch Size and Code Modifications in Incomplete Fixes}
Table~\ref{tab:patch-stats} summarizes code-level characteristics of the
collected patches.  Commit messages are excluded as they do not reflect actual
code modifications.  We use \texttt{diffstat} to compute added, deleted, and
modified lines of code (mean/std).

\textbf{Finding 6:}
As shown in Table~\ref{tab:patch-stats}, most patches involve only modest code
modifications.
Only a few patches modify substantial amounts of code.
For example, the fix for CVE-2016-6786 adds 207 lines and deletes 37 lines.
Across multiple dimensions, Fix-1 patches tend to be smaller than Fix-0 patches.
Overall, incomplete security fixes typically involve patches of limited
complexity, facilitating root-cause analysis.

\section{RQ2: What Are the Root Causes of Incomplete Security Bug Fixes?}
\label{sec:root-cause}

In this section, we analyze the root causes of the \ncases{} incomplete security
bug fixes collected from the Linux kernel.
We categorize cases into three phases before fixing, during fixing, and after
fixing, highlighting representative examples and summarizing common patterns.

\begin{figure}[!t]
\centering
\includegraphics[width=\linewidth]{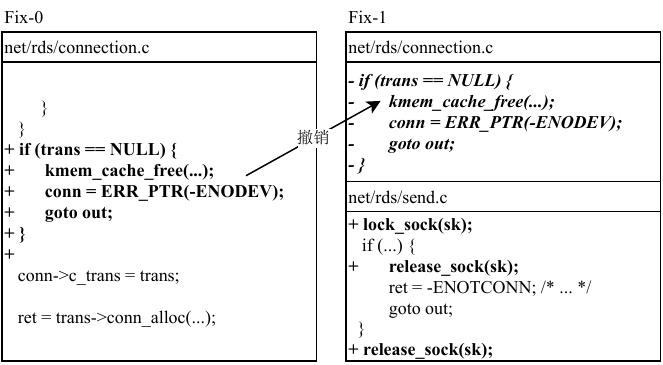}
\caption{CVE-2015-6927 (left) and CVE-2015-7990 (right)}
\label{fig:cve-2015-7990}
\end{figure}

\begin{figure}[!t]
\centering
\includegraphics[width=\linewidth]{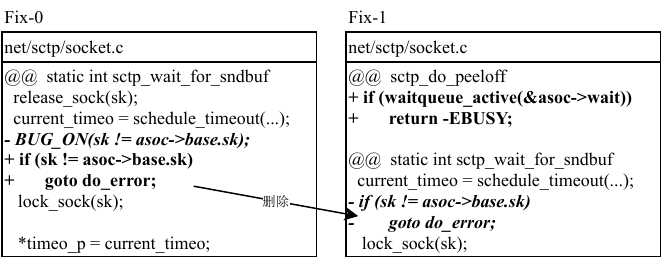}
\caption{CVE-2017-5986 (left) and CVE-2017-6353 (right)}
\label{fig:cve-2017-6353}
\end{figure}

\subsection{Misled by Symptoms Before Fixing}

This category highlights that developers were misled by crash symptoms and
failed to identify the underlying root cause.

\begin{itemize}
\item \textbf{CVE-2015-6937 $\rightarrow$ CVE-2015-7990:}  

In Figure~\ref{fig:cve-2015-7990}, Fix-0 attempted to resolve a NULL-pointer
dereference but did so incorrectly.
Fix-1 correctly resolves the issue by introducing appropriate locking.

\item \textbf{CVE-2017-5986 $\rightarrow$ CVE-2017-6353:}  

As shown in Figure~\ref{fig:cve-2017-6353}, Fix-0 removed an assertion
improperly, leading to a double-free vulnerability.
Fix-1 applies a new fixing strategy that avoids both issues and correctly
resolves the flaw.

\end{itemize}

\noindent
\textbf{Takeaway 1:}
Before applying a fix, developers must thoroughly understand the root cause
rather than rely solely on surface-level crash symptoms.
Such cases typically require reverting Fix-0 and introducing a new, corrected
fix.

\begin{figure}[!t]
\centering
\includegraphics[width=\linewidth]{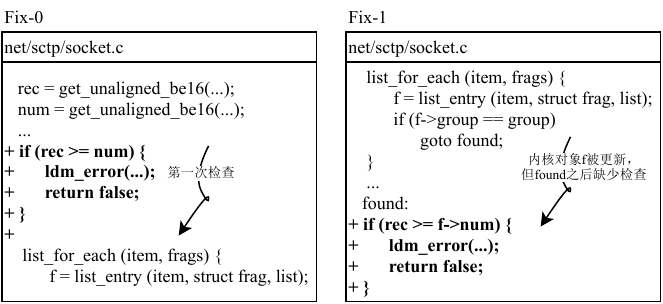}
\caption{CVE-2011-1017 (left) and CVE-2011-2182 (right)}
\label{fig:cve-2011-2182}
\end{figure}

\begin{figure}[!t]
\centering
\includegraphics[width=\linewidth]{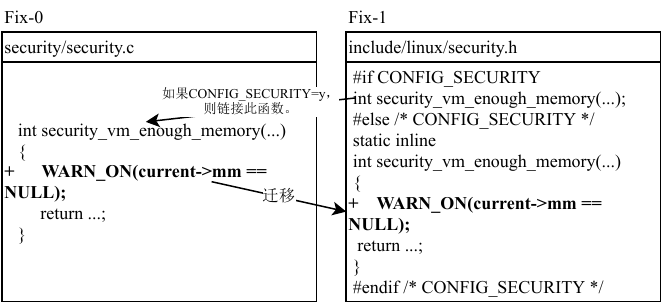}
\caption{CVE-2010-1643 (left) and CVE-2008-7256 (right)}
\label{fig:cve-2008-7256}
\end{figure}

\begin{figure}[!t]
\centering
\includegraphics[width=\linewidth]{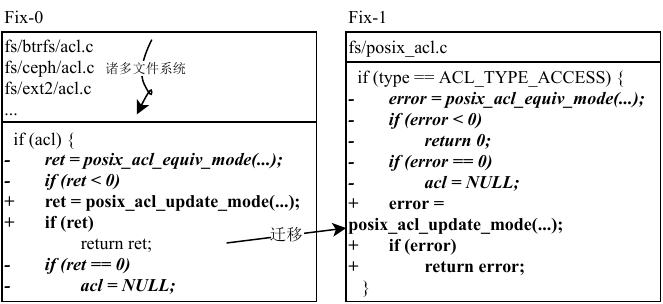}
\caption{CVE-2016-7097 (left) and CVE-2017-5551 (right)}
\label{fig:cve-2017-5551}
\end{figure}

\subsection{Overlooking Similar Issues During Fixing}

This category shows that developers correctly identified the root cause but
missed analogous issues in similar modules.

\begin{itemize}

\item \textbf{CVE-2011-1017 $\rightarrow$ CVE-2011-2182:}  
As shown in Figure~\ref{fig:cve-2011-2182}, Fix-0 added a check on \texttt{rec},
but this single check was insufficient.
Fix-1 completes the fix by adding the missing condition.

\item \textbf{CVE-2010-1643 $\rightarrow$ CVE-2008-7256:}
As shown in Figure~\ref{fig:cve-2008-7256}, Fix-0 patched
\texttt{security\_vm\_enough\_memory()}, but failed to modify the corresponding
function guarded by \texttt{CONFIG\_SECURITY}.
Fix-1 resolves this oversight.

\item \textbf{CVE-2016-7097 $\rightarrow$ CVE-2017-5551:}  
As shown in Figure~\ref{fig:cve-2017-5551}, Fix-0 addressed the issue in
multiple filesystems, but omitted the tmpfs variant.
Fix-1 completes the fix.

\item \textbf{CVE-2019-14615 $\rightarrow$ CVE-2020-8832:}  
The original vulnerability had been fixed upstream, but the patch was not back-ported to a specific Ubuntu release.
\end{itemize}

\noindent
\textbf{Takeaway 2:}
When fixing a vulnerability, developers should check whether analogous modules
contain similar issues, e.g., different call sites, similar functions, parallel
subsystems (filesystems, network protocols, CPU architectures), different
branches, or downstream distributions.
These cases often require migrating Fix-0 logic to corresponding modules.

\begin{figure}[!t]
\centering
\includegraphics[width=\linewidth]{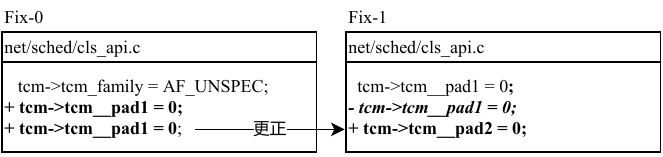}
\caption{CVE-2005-4881 (left) and CVE-2009-3612 (right)}
\label{fig:cve-2009-3612}
\end{figure}

\subsection{Introducing Semantic Errors After Fixing}

This category shows that developers understood the root cause but introduced new
semantic errors during the fix.

\subsubsection{Variable Level}

\begin{itemize}
\item \textbf{CVE-2005-4881 $\rightarrow$ CVE-2009-3612:}  
As shown in Figure~\ref{fig:cve-2009-3612}, Fix-0 mistakenly reinitialized
\texttt{tem\_pad1}, leaving \texttt{tem\_pad2} uninitialized and causing an
information leak.  Fix-1 initializes \texttt{tem\_pad2} properly.
\end{itemize}

\begin{figure}[!t]
\centering
\includegraphics[width=\linewidth]{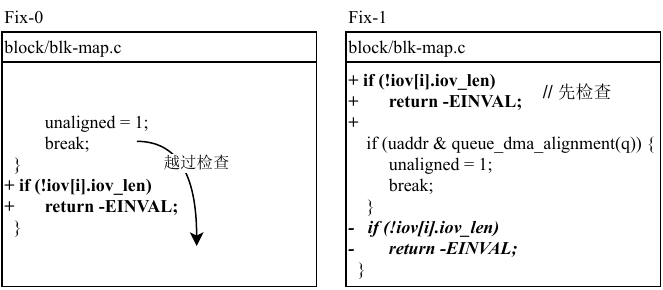}
\caption{CVE-2010-4163 (left) and CVE-2010-4668 (right)}
\label{fig:cve-2010-4668}
\end{figure}

\begin{figure}[!t]
\centering
\includegraphics[width=\linewidth]{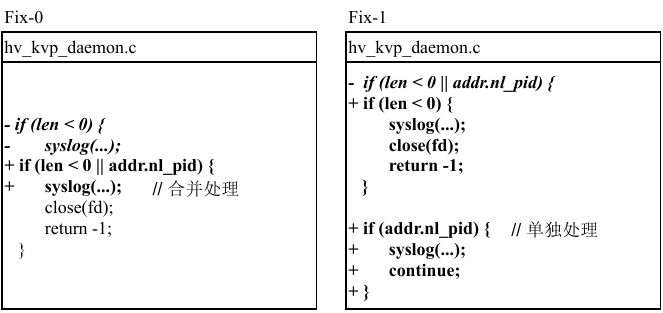}
\caption{CVE-2012-2669 (left) and CVE-2012-5532 (right)}
\label{fig:cve-2012-5532}
\end{figure}

\begin{figure}[!t]
\centering
\includegraphics[width=\linewidth]{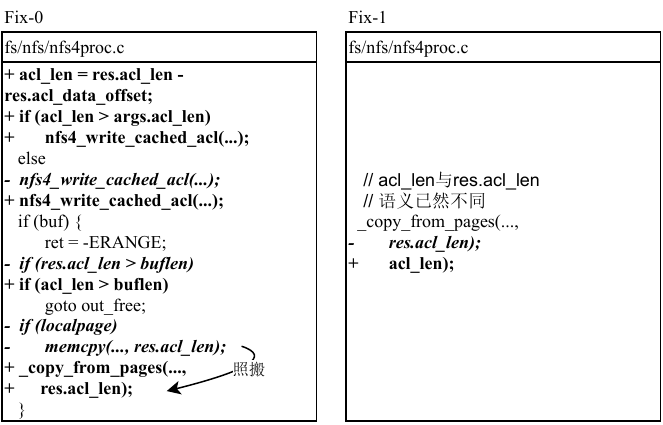}
\caption{CVE-2011-4131 (left) and CVE-2012-2375 (right)}
\label{fig:cve-2012-2375}
\end{figure}

\begin{figure}[!t]
\centering
\includegraphics[width=\linewidth]{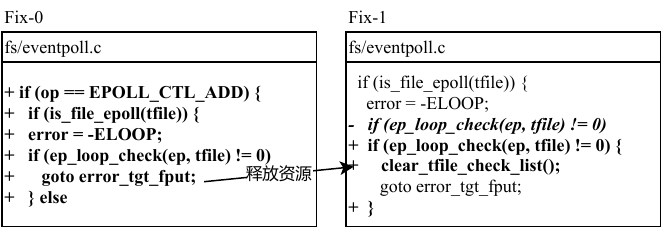}
\caption{CVE-2011-1083 (left) and CVE-2012-3375 (right)}
\label{fig:cve-2012-3375}
\end{figure}

\begin{figure*}[!t]
\centering
\includegraphics[width=\textwidth]{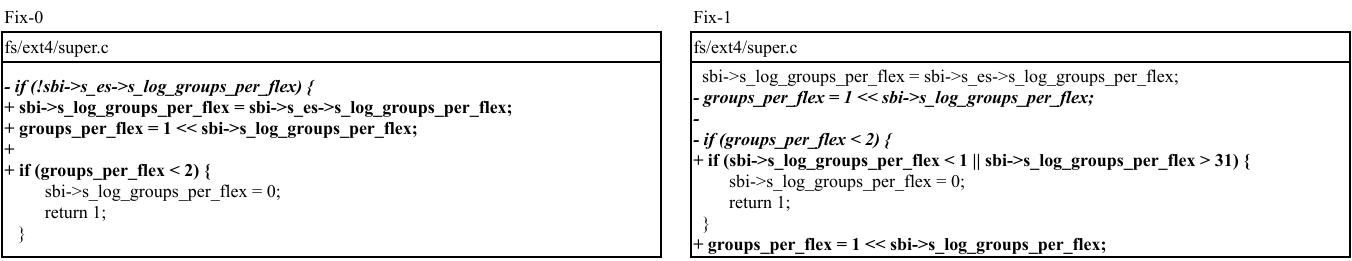}
\caption{CVE-2009-4307 (left) and CVE-2012-2100 (right)}
\label{fig:cve-2012-2100}
\end{figure*}

\subsubsection{Function Level}

\begin{itemize}
\item \textbf{CVE-2010-4163 $\rightarrow$ CVE-2010-4668:}  
As shown in Figure~\ref{fig:cve-2010-4668}, Fix-0 introduced a check on
\texttt{iov[i].iov\_len}, but placed it incorrectly such that the check was
bypassed when the \texttt{break} was triggered.
Fix-1 moves the check before the break to preserve intended semantics.

\item \textbf{CVE-2012-2669 $\rightarrow$ CVE-2012-5532:}  
In Figure~\ref{fig:cve-2012-5532}, Fix-0 conflated the cases of
\texttt{addr.nl\_pid != 0} and \texttt{len < 0}, violating semantic expectations
and introducing new issues.
Fix-1 handles \texttt{addr.nl\_pid != 0} separately.

\item \textbf{CVE-2011-4131 $\rightarrow$ CVE-2012-2375:}  
In Figure~\ref{fig:cve-2012-2375}, Fix-0 incorrectly reused the last parameter passed to \texttt{memcpy}, despite differing semantics between \texttt{res.acl\_len} and \texttt{acl\_len}.  
Fix-1 corrects this mismatch.

\item \textbf{CVE-2009-4307 $\rightarrow$ CVE-2012-2100:}  
As shown in Figure~\ref{fig:cve-2012-2100}, Fix-0 attempted to validate
\texttt{groups\_per\_flex} but left overflow cases unhandled. 
Fix-1 correctly validates \texttt{sbi->s\_log\_groups\_per\_flex}. 

\end{itemize}

\begin{figure*}[!t]
  \centering
  \includegraphics[width=0.75\textwidth]{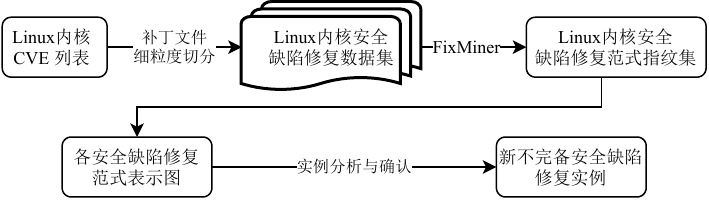}
  \caption{Workflow of the fix-pattern–fingerprinting detection algorithm}
  \label{fig:algo} 
\end{figure*}

\begin{figure*}[!t]
\centering
\includegraphics[width=.8\textwidth]{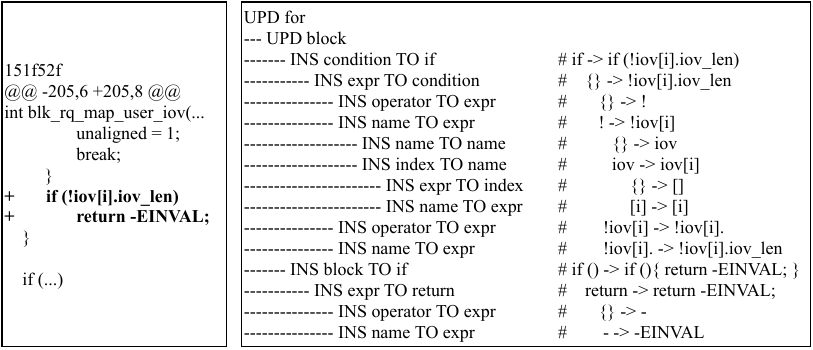}
\caption{Example of a fix pattern: the patch (left) and the corresponding
FixMiner-generated fix pattern (right)}
\label{fig:fix-pattern}
\end{figure*}

\begin{figure*}[t]
\centering
\includegraphics{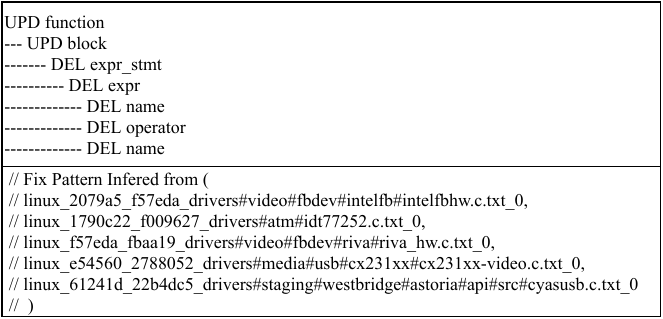}
\caption{Example fix-pattern fingerprint shared among multiple patches}
\label{fig:fix-pattern-fingerprinting}
\end{figure*}

\begin{figure*}[t]
\centering
\includegraphics[width=.8\textwidth]{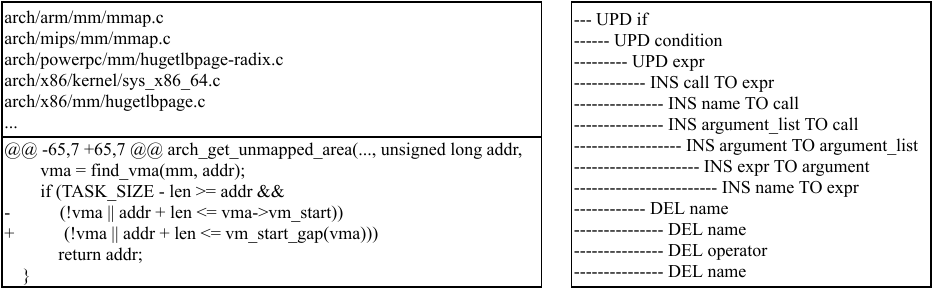}
\caption{Fix pattern of CVE-2017-1000365}
\label{fig:cve-2017-1000365-pattern}
\end{figure*}

\subsubsection{Resource Level}

\begin{itemize}
\item \textbf{CVE-2011-1083 $\rightarrow$ CVE-2012-3375:}  
As shown in Figure~\ref{fig:cve-2012-3375}, Fix-0 introduced new data structures
but failed to release all allocated resources, leading to a DoS vulnerability.  
Fix-1 releases the missing resources.

\item \textbf{CVE-2013-4312 $\rightarrow$ CVE-2016-2550:}  
Fix-0 failed to maintain the reference count for open files, enabling an attacker to exhaust memory by repeatedly opening files.

\item \textbf{CVE-2010-4250 $\rightarrow$ CVE-2011-1479:}  
Fix-0 was too simplistic and introduced a double-free vulnerability.
\end{itemize}

\subsubsection{Access-Control Level}

\begin{itemize}

\item \textbf{CVE-2010-4347 $\rightarrow$ CVE-2011-1021:}  
Fix-0 reduced excessive write permissions of \texttt{custom\_method} in debugfs,
but mistakenly preserved write access for the root user.  
Fix-1 restricts access by enabling the file only in debugging mode.
Similarly, when addressing CVE-2016-9576, the fix failed to consider the case
where the KERNEL\_DS segment selector is active. Under this condition, an
attacker could exploit sg\_write to gain arbitrary read/write primitives or
trigger a denial-of-service attack (CVE-2016-10088).

\end{itemize}

\noindent
\textbf{Takeaway 3:}
After fixing a vulnerability, developers, testers, and reviewers must verify
that the fix preserves intended program semantics.
Such cases typically require analyzing at which granularity Fix-0 violated the
original semantics and applying a targeted correction.

\section{RQ3: Detecting Incomplete Security Bug Fixing due to ``Missing
Similar Components''}
\label{sec:detection}

In Section~\ref{sec:root-cause}, we analyzed the phenomenon of incomplete
security bug fixing in the Linux kernel and summarized common mistakes across
the three stages of fixing (before, during, and after). 
Based on the insights drawn from the category of ``missing similar components
during fixing'', this section presents a detection algorithm based on
\emph{fix-pattern fingerprints} to automatically identify other potential
incomplete security bug fixes in the Linux kernel.
This method helps reveal hidden incomplete fixes and can reduce the likelihood
of such issues during patch submission.

\subsection{Fix-Pattern–Fingerprint Detection}

Incomplete security bug fixes of the ``missing similar components'' type can be
characterized by the following rule: 
Within a set of similar components, the Fix-0 patch updates only a subset of
them while omitting others.
Here, a component is an abstract concept that may refer to a group of similar
functions or a group of modules implementing comparable functionality.
In the patch file submitted for a security bug fix, if identical or similar fix
patterns are applied to multiple similar components, then the complete set of
such components can be further examined to check whether any have been omitted.
Following Fig.~\ref{fig:algo}, we applied FixMiner~\cite{anil2020fixminer} to
extract and analyze fix patterns from 1,615 CVE fixes in the Linux kernel and
ultimately uncovered two new problematic cases.

We collected 1,615 Linux kernel CVEs using the Linux Kernel CVE
database~\cite{linuxkernelcves}, each of which includes commit hashes
referencing its fixes.
We then constructed a dataset for fix-pattern analysis.
For each CVE we retrieved all associated patch files, as well as snapshots of
the patched file before and after the patch.
If a single CVE patch modified multiple files, we split it into several
fine-grained patch files, each corresponding to one modified source file,
resulting in 2,753 fine-grained patches, facilitating detailed analysis.

FixMiner was then applied to this dataset to extract fix-pattern fingerprints.
FixMiner, built on top of GumTreeDiff~\cite{gumtreediff}, automatically extracts
abstract syntax tree (AST)–based fix patterns by abstracting the code changes
within patches. 
Figure~\ref{fig:fix-pattern} shows an example patch and its corresponding fix
pattern.
After configuration and processing, FixMiner generated fix-pattern fingerprints
for all 2,753 fine-grained patches.
It also computed similarities among fix patterns and linked them accordingly.
Ultimately, FixMiner identified 311 distinct fix-pattern fingerprints related to
security bug fixes.  Figure~\ref{fig:fix-pattern-fingerprinting} shows an
example fingerprint shared across multiple patches.

Using these fingerprints, we constructed a fix-pattern graph for each CVE. 
The root node represents the CVE ID; child nodes correspond to fix-pattern
fingerprints that are detected in Fix-0 across its modified files. 
A large number of repeated nodes within a graph indicates that the Fix-0 patch
likely involved modifications across many similar components. 
We then conducted manual validation for these candidates.

\subsection{New Findings}

\textbf{Finding 7:}  
From scanning the patches of 1615 CVEs, we identified two incomplete security bug fixes.  
One has already been confirmed by the Linux kernel community and is in the process of being merged into the mainline;  
the other was a hidden incomplete fix that had been silently corrected in a later patch without receiving a CVE ID.

\begin{itemize}
\item \textbf{CVE-2017-1000364:}  
CVE-2017-1000364 was submitted in June 2017.  
Its fix pattern, illustrated in Fig.~\ref{fig:cve-2017-1000365-pattern}, was
applied to multiple architectures—Arc, ARM, FRV, MIPS, Parisc, PowerPC, s390,
SH, Sparc, Tile, x86, and Xtensa.  
During manual analysis, we found that the fix pattern was \emph{not} applied to
the NDS32 architecture.
Support for NDS32 was added to the Linux kernel in 2018.
The newly added module reused existing code to implement standard kernel
features such as memory mapping.
However, the developers failed to port the CVE-2017-1000364 fix to NDS32 when
integrating the new architecture.
As a result, the vulnerability described in the CVE persisted on NDS32, even
though all other architectures were properly patched.

\begin{figure}[!t]
\centering
\includegraphics[width=0.6\linewidth]{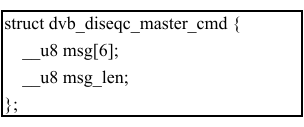}
\caption{Definition of DiSEqC messages}
\label{fig:disqec-definition}
\end{figure}

\begin{figure}[!t]
\centering
\includegraphics[width=0.6\linewidth]{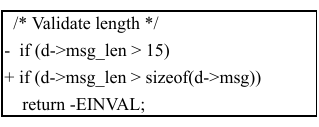}
\caption{Fix pattern of CVE-2015-9289}
\label{fig:disqec-fixing}
\end{figure}

\begin{table*}[!t]
\centering
\caption{Comparison of Studies on Root Cause Analysis of Bug Fixing}
\label{tab:papers}
\footnotesize
\begin{tabular}{@{}cccc@{}}
\toprule
Study & Dataset Source (System Software) & Focus & Detection Tool \\
\midrule
BuJihun Park et al.~\cite{r1-c-msr12} & Eclipse and Mozilla (No) & Incomplete General Bug Fixes & \ding{55} \\
Le An et al.~\cite{an2014supplementary} & Eclipse and Mozilla, etc. (No) & Incomplete and Re-opened General Bug Fixes & \ding{51} \\
Qing Mi et al.~\cite{mi2016empirical} & Eclipse (No) & Re-opened General Bug Fixes & \ding{55} \\
Zuoning Yin et al.~\cite{yin2011fixes} & Linux, FreeBSD kernel, etc. (Yes) & Incomplete General Bug Fixes & \ding{55} \\
Zhen Ni et al.~\cite{ni2020analyzing} & Radare2 and Mozilla (No) & General Bug Fixes & \ding{55} \\
Kai Zhang et al.~\cite{mozilla2017} & Mozilla (No) & Incomplete Security Bug Fixes & \ding{55} \\
\midrule
This Work & \textbf{Linux Kernel (Yes)} & \textbf{Incomplete Security Bug Fixes} & \ding{51} \\
\bottomrule
\end{tabular}
\end{table*}

\begin{figure}[t]
\centering
\includegraphics[width=\linewidth]{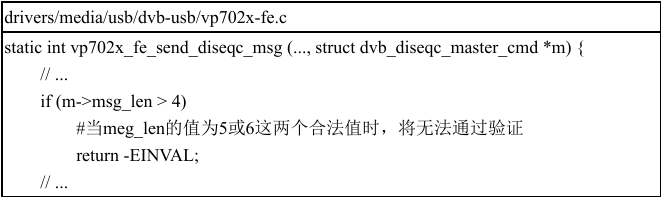}
\caption{Incorrect checking of DiSEqC message length}
\label{fig:disqec-wrong-fix}
\end{figure}

\item \textbf{CVE-2015-9289:}
This CVE addressed a buffer overflow in multimedia DVB drivers in the Linux
kernel.  
The DiSEqC message structure is shown in Fig.~\ref{fig:disqec-definition}.  
The field \textbf{msg} holds the message data, and \textbf{msg\_len} indicates
its length. 
A valid DiSEqC message ranges from 3 to 6 bytes (3-byte header plus
up to 3 bytes of optional parameters).
The CVE patch corrected improper upper-bound checks on \textbf{msg\_len} in
several DVB drivers (Fig.~\ref{fig:disqec-fixing}).  
However, our analysis found that several DVB drivers in the same subsystem still
contained incorrect checks (Fig.~\ref{fig:disqec-wrong-fix}).
Some drivers (e.g., \texttt{tda8083.c}, \texttt{stv0288.c}, \texttt{stv0299.c},
\texttt{cx24123.c}, \texttt{ds3000.c}) lacked \emph{any} length check.
Thus, the patch for CVE-2015-9289 was incomplete.  
Interestingly, maintainers later fixed these missing checks through a normal
bug-fix patch without applying for a new CVE ID.
This hidden incomplete fix expands our dataset with another overlooked instance.

\end{itemize}

\section{Related Work}\label{sec:related-work}

\subsection{Incomplete Security Bug Fixes}

Bug fixing is a common yet essential activity throughout the software
development lifecycle.
Over the past decade, numerous studies have examined the characteristics of
general bug
fixes~\cite{r1-c-msr12,an2014supplementary,r1-c-icse15,bohme2017bug,yue2017characterization,wang2018empirical,rwemalika2019industrial,wang2020empirical}
and security bug
fixes~\cite{r1-v-icse12,r1-v-ccs17,r1-v-usenix18,morrison2018vulnerabilities,piantadosi2019fixing,wang2019characterizing,canfora2020investigating},
with the goals of improving bug management, refining fixing practices, and
facilitating automated bug-fixing approaches.
Most
studies~\cite{r1-v-ccs17,r1-c-icse15,bohme2017bug,yue2017characterization,wang2018empirical,r1-v-icse12,morrison2018vulnerabilities,piantadosi2019fixing,wang2019characterizing,canfora2020investigating}
focus on answering a set of fundamental questions: the general workflow of bug
fixing; whether fixes exhibit spatial locality or repetition at the levels of
lines, functions, and files; the extent to which fixes rely on API semantics;
required techniques, human effort, and time for locating, diagnosing, and fixing
bugs; differences between general and security bug fixes; and the potential for
automated bug fixing.
Others explore the number of patches involved in bug fixing and its
implications.

Bug fixes can be broadly divided into two categories depending on the number of
patches involved.  The first category involves a single patch; the second
involves multiple patches and is referred to as an \emph{incomplete fix}.
Generally, incomplete fixes fall into two subtypes: \emph{incomplete fixes} and
\emph{incorrect fixes}.  An incomplete fix occurs when the initial patch does
not repair all code locations affected by the flaw.  An incorrect fix refers to
a case where the initial patch is flawed—failing to remediate the original issue
or introducing new defects.  Meanwhile, several studies have highlighted
significant differences between the fixing processes of general bugs and
security bugs, advocating for treating the two categories
separately~\cite{zaman2011security,camilo2015bugs,canfora2020investigating}.
Existing studies of incomplete fixes over the past decade are limited in number
and mostly focus on non-system software such as Eclipse and
Mozilla~\cite{r1-c-msr12,an2014supplementary,mi2016empirical,mozilla2017}.
Notably, Yin et al.~\cite{yin2011fixes} examined incomplete fixes in the Linux
kernel in 2011, but their study analyzed both general and security bug fixes
jointly, emphasizing the importance of understanding incomplete fixes,
summarizing fixing patterns and common mistakes, and providing general
recommendations.  However, their dataset lacks the past ten years of security
bug fixes and is thus outdated.

\textbf{As shown in Table~\ref{tab:papers}, our work differs by focusing
specifically on incomplete \emph{security} bug fixes in the Linux kernel,
including cases from the \textbf{last decade}, providing novel insights.
Moreover, building upon these findings, we develop a \textbf{detection tool}
targeting a prevalent class of incomplete security bug fixes, thereby improving
the correctness of security patches and advancing the state of the art.}

\subsection{Root Cause Analysis of Software Bugs}

Several studies over the past decade have investigated the root causes of
software
bugs~\cite{yin2011fixes,r1-c-msr12,an2014supplementary,mi2016empirical,ni2020analyzing,mozilla2017}.
Park et al.~\cite{r1-c-msr12} analyzed incomplete general bug fixes in Eclipse
JDT Core, Eclipse SWT, and Mozilla, identifying diverse causes such as forgotten
code porting, incorrect handling of conditional branches, and incomplete
refactoring.  They emphasized that similarity-based code detection is
insufficient to prevent all incomplete fixes.  They also found that in 14\%–15\%
of cases, subsequent patches differed substantially from the initial patch both
in location and structural dependency, implying that automated patch
recommendation systems need redesign.  Le An et al.~\cite{an2014supplementary}
studied the relationship between incomplete bug fixes and re-opened bugs across
Mozilla, NetBeans, Eclipse JDT Core, Eclipse Platform SWT, and WebKit.  They
found partial overlap between the two phenomena and further developed
classifiers based on human factors, bug reports, and simple fixing statistics to
distinguish them.
Zhang et al.~\cite{mozilla2017} examined incomplete security bug fixes in
Mozilla and found that such fixes correlate with the complexity of root causes
and complexity of the fix itself; they concluded that more elaborate and
effective fixing practices can reduce the likelihood of incomplete security
fixes.
Our work, in contrast, focuses on incomplete security fixes specifically within
the Linux kernel, a large-scale system software project.

\subsection{Similarity-Based Bug Detection}

Leveraging known fix
patterns~\cite{r1-c-msr16,campos2017common,liu2018mining,anil2020fixminer},
various similarity-based techniques have been proposed over the past decade to
detect software bugs in open-source projects~\cite{zhong2020inferring}.  Zhong
et al.~\cite{zhong2020inferring} used the Grapa tool to analyze code changes
before and after bug fixes, extracting bug features from patches and building a
classification model.  They conducted the first large-scale evaluation on 6,048
bug fixes from four popular Apache projects, confirming three new defects, all
of which were subsequently acknowledged and fixed.  Our work adopts a similar
philosophy but differs in purpose and application: we use fix-pattern
fingerprints extracted from incomplete \emph{security} bug fixes to detect
additional incomplete security bug fixes in the Linux kernel.  By constructing a
dataset of 1,615 CVE-related Linux kernel security fixes and extracting
fix-pattern fingerprints, we successfully identified one previously unknown and
unfixed incomplete security fix, as well as one hidden but already-fixed case.

\section{Limitations and Future Directions}\label{sec:discussion}

This paper presents the characteristics, root causes, and detection techniques
of incomplete security bug fixes in the Linux kernel. However, several
limitations remain.
(1) In analyzing the characteristics and root causes of incomplete security bug
fixes in the Linux kernel, we relied primarily on manual analysis due to the
relatively small dataset. Manual analysis is inefficient when scaling to all
incomplete security bug fixes across large-scale open-source software
ecosystems. In the future, improvements can be made from two aspects of static
program analysis on patches.  First, during our study, we found that patch files
and their commit messages often diverge. An automated technique that summarizes
patch semantics and contrasts them with the accompanying descriptions would
greatly aid in understanding, testing, and auditing security fixes.  Second,
although Fix-1 patches reflect the presence of incomplete security bug fixes, it
is often difficult to directly identify their underlying causes without further
analysis. If the manual reasoning process can be formalized and automated, it
would significantly accelerate the entire workflow.
(2) In detecting incomplete security bug fixes in the Linux kernel, this paper
focuses only on the ``missing similar components'' category. Other categories
lack clear patterns or exhibit high complexity, making them more challenging to
detect. Future research may proceed in two directions.
First, by studying incomplete security bug fixes in a wider range of open-source
projects, more root causes may be identified, enabling the design of new
detection algorithms. 
Second, for specific classes of vulnerabilities, e.g., data races, highly
specialized detection algorithms may be developed to improve coverage.

\section{Conclusion}\label{sec:conclusion}

This paper presents the first dataset of incomplete security bug fixes in the
Linux kernel, and provides a comprehensive analysis of their characteristics,
root causes, and detection techniques. Finally, by applying our
fix-pattern–based fingerprinting algorithm to 1,615 Linux kernel
vulnerabilities, we discovered one previously unnoticed incomplete security fix
and one newly identified incomplete security fix. These findings demonstrate the
potential of automated techniques to aid in the detection of incomplete security
bug fixes and thereby contribute to enhancing the security of the Linux kernel.

\bibliographystyle{plain}
\bibliography{main.bib}

\end{document}